\documentclass[twocolumn]{aastex631}

\usepackage{amsmath}
\usepackage{bm}
\usepackage{color}
\usepackage{multirow}

\newcommand{\psr}{B0818$-$41}
\newcommand{\phase}{\phi}
\newcommand{\Nsp}{\ensuremath{n}}
\newcommand{\vref}{\ensuremath{\nu_\text{ref}}}
\newcommand{\Dphs}{\Delta\phase_s}
\newcommand{\Dphc}{\Delta\phase_c}
\newcommand{\DphDR}{\Delta\phase_\text{DR}}

\newcommand{\DM}{\text{DM}}
\newcommand{\DMsym}{\ensuremath{\text{DM}_\text{sym}}}
\newcommand{\DDM}{\Delta\DM}
\newcommand{\phDM}{\phase_{\DM}}

\newcommand{\DMunits}{pc$\,$cm$^{-3}$}

\newcommand{\deriv}[2]{\frac{\text{d}#1}{\text{d}#2}}
\newcommand{\Deriv}[2]{\dfrac{\text{d}#1}{\text{d}#2}}
\newcommand{\dd}[2]{\frac{\text{d}^2{#1}}{\text{d}{#2}^2}}

\newcommand{\sAl}{\sin\alpha}
\newcommand{\cAl}{\cos\alpha}

\newcommand{\sZe}{\sin\zeta}
\newcommand{\cZe}{\cos\zeta}

\newcommand{\tPs}{\tan\psi}

\begin{document}
\title{Testing the circularity of PSR \psr{}'s carousel}

\author[0000-0001-6114-7469]{Samuel J. McSweeney}
\affiliation{International Centre for Radio Astronomy Research, Curtin University, Bentley, WA 6102, Australia}

\author{Lisa Smith}
\affiliation{International Centre for Radio Astronomy Research, Curtin University, Bentley, WA 6102, Australia}

\author[0000-0002-8383-5059]{N. D. Ramesh Bhat}
\affiliation{International Centre for Radio Astronomy Research, Curtin University, Bentley, WA 6102, Australia}

\author[0000-0001-8193-0557]{Geoff Wright}
\affiliation{Jodrell Bank Centre for Astrophysics, School of Physics and Astronomy, University of Manchester, M13 9PL, UK}

\begin{abstract}
    The phenomenon of sub-pulse drifting is an important single-pulse phenomenon that can potentially provide important insights into the elusive radio emission mechanism in pulsars.
    We analyze the frequency behaviour of the single pulses of \psr{}, observed from $300$ to $500\,$MHz (Band 3 of the uGMRT), and compare it to the evolution of the average profile to place constraints on the geometry of the pulsar's emission beam.
    We show that a circular carousel of discrete beamlets, where each beamlet has radial symmetry, is not consistent with the observed behaviour, and describe an alternative, consistent range of possible elliptical carousel geometries.
    We also combine the uGMRT data with some archival MWA observations and several other published profiles to characterize the profile evolution across a frequency range spanning ${\sim}170$ MHz to ${\sim}1.4\,$GHz.
\end{abstract}


\section{Introduction} \label{sec:introduction}

The physical mechanism that gives rise to neutron stars' pulsed radio emission is still poorly understood, despite more than half a century of dedicated investigation.
On the theoretical side, the difficulty stems from the uniqueness of the neutron star (or pulsar) environment, our limited understanding of relativistic plasmas, and our inability to reproduce the enormous gravitational and magnetic fields to conduct controlled experiments on Earth \citep{Melrose2017}.
On the observational side, the difficulty is rooted in the bewildering array of observed morphological features and emission patterns, which differ from pulsar to pulsar, and which any candidate emission mechanism must be able to account for.
Consequently, many studies focus on individual pulsars, with the hope that the successful detailed modelling of one pulsar may have features that generalize to other pulsars.

In this paper, we continue this program and revisit the feature-rich emission of PSR \psr{} using archival observations in Band 3 (300 to 500 MHz) of the upgraded Giant Metrewave Radio Telescope \citep[uGMRT;][]{Reddy2017}.
The combination of the uGMRT's sensitivity and wideband coverage provides a superb view of \psr{}'s single pulses, allowing two of its most striking features to be studied simultaneously in unprecedented detail: its sub-pulse drifting and profile evolution.
Both of these phenomena have individually been studied in \psr{} extensively (as outlined below) but here we present the first analysis of the frequency evolution of its individual pulses.
In the context of the carousel model \citep{Ruderman1975,Deshpande2001,Edwards2002}, which we adopt here, this allows us to place new constraints on the geometry and dynamics of \psr{}'s emission beam.

\begin{figure}[th]
    \centering
    \includegraphics[width=1.0\linewidth]{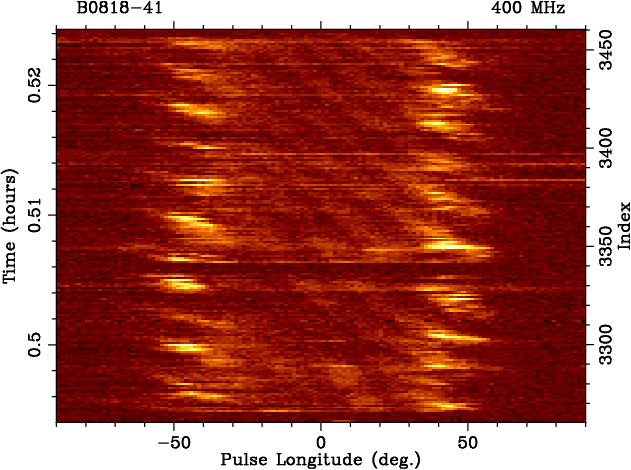}
    \caption{A pulsestack of ${\sim}200$ pulses produced using the uGMRT data set (see \S\ref{sec:observations} for details), showing the drifting behaviour in both the outer and inner regions. A few nulling pulses are visible at the top and bottom of the pulsestack, as well as a few in the middle.}
    \label{fig:gmrt_pulsestack2}
\end{figure}

Much work has already been done to characterize the sub-pulse drifting behavior of \psr{} in terms of the carousel model, primarily in the series of papers by \citet{Bhattacharyya2007,Bhattacharyya2009,Bhattacharyya2010a}.
They showed that the sub-pulse drifting occurs across the whole pulse window, but that the drift bands associated with the profile's prominent outer components (see Fig. \ref{fig:gmrt_pulsestack2}) are distinct from those associated with the inner components, or bridge emission.
The drift bands in the two regions (outer vs inner) have a different drift rate but the same value of $P_3 \approx 18.3\,P$, which is the time it takes for a sub-pulse to reappear at the same rotation phase, where $P = 0.545\,$s is the rotation period.
Having the same $P_3$ value but different drift rates implies that the time between consecutive sub-pulses, $P_2$ is also different in the two regions.

Taking all these features together, they modelled the drifting behavior as arising from the line of sight (LoS) cutting through two concentric, nested, phase-locked carousels, reminiscent of the nested hollow cone models of \citet{Rankin1983,Rankin1993}.
\citet{Bhattacharyya2009a} estimate the number of sparks in each carousel to be ${\sim}20$, with the whole carousel making one complete revolution once every $P_4 \approx 18\,P$.
They found two viewing geometries which were consistent with the observed polarization angle (PA) sweep according to the rotating vector model \citep[RVM;][]{Radhakrishnan1969}, and also consistent with the drifting behavior of the sub-pulses.
As can be expected for wide-profile pulsars like \psr{}, both viewing geometries are ``nearly-aligned,'' with small magnetic inclination $\alpha$ and impact angle $\beta$.

In the solutions of \citet{Bhattacharyya2009a}, the carousel rotation rate is aliased with the stellar rotation, in the sense that the time it takes for a spark to reach the instantaneous position of its predecessor, $\hat{P}_3 = P_4/\Nsp$, where $\Nsp$ is the number of sparks in the carousel, is similar to the rotation rate.
The observed $P_3$ is just an aliased form of $\hat{P}_3$, the more intrinsic and physically meaningful property.
The assumed presence of aliasing has the added benefit of bringing the carousel rotation rate more in line with the original prediction of \citet{Ruderman1975}, which for \psr{} is $P_4 \approx 2\,P$ \citep[cf. Eq. (7) in][]{McSweeney2019a}.
Alternatively, \citet{Morozova2014} argue that the revized sub-pulse drift model of \cite{Leeuwen2012}, which predicts a much slower $P_4 \approx 377\,P$ for \psr{}, is more in line with the unaliased rate of $P_4 \approx 343\,P$.

\citet{Bhattacharyya2009a} also analyze the profile evolution across five frequencies between $157\,$MHz and $1060\,$MHz.
They state that the separation between the peaks of the outer components goes as $\Delta\phase \propto \nu^{\alpha}$ with $\alpha = -0.2$, but a least squares fit to their measured peak separations gives a value closer to $\alpha \approx -0.27$.
As they point out, the frequency evolution of \psr{} is typical---the component separation increases towards lower frequencies---and entirely consistent with received ideas of a radius-to-frequency mapping \cite[RFM;][]{Cordes1978} being the root cause.

With sufficiently bright and wideband observations in hand, we can start to probe some of the more subtle effects expected for the carousel model.
One such effect was explored in great detail in B0943+10 by \citet{Bilous2018}, in which the finite size and shape of the discrete emission beam components (hereafter called \textit{beamlets}) affect the phase at which the corresponding driftbands appear at different frequencies.
Assuming that the sole effect of the RFM is simply to move the beamlets radially with respect to the magnetic axis, carousels of different sizes (i.e. observed at different frequencies) naturally intersect the LoS traverse at different rotation longitudes.

In addition to the RFM-induced phase shift just described, there is a secondary effect relating to the extra time it takes a given beamlet to ``catch up'' to the intersection point.
In her analysis, \citet{Bilous2018} folded the pulsestacks modulo $P_3$ (so-called \textit{modfolds}) to track how the centroid positions of the driftbands move both in phase and in time (i.e. at later pulses) as a function of observing frequency across the large fractional bandwidth of her Low Frequency Array (LOFAR) low-band observations.
The movement of the driftbands, she showed, could be well described by, and is a natural consequence of, the carousel model.

In principle, the same analysis can be applied to any sub-pulse drifting pulsar whose drifting behavior is steady enough to make modfolds.
\psr{} almost qualifies: its otherwise stable drifting pattern is frequently interrupted by nulling sequences (where the emission ceases entirely), which are associated with disturbances of the drift rate in the surrounding burst sequence \citep{Bhattacharyya2010a}.
Although it might be possible to circumvent these issues by selecting only pulses in stable drifting sequences, in this paper we adopt a different approach which considers how the arguments of \citet{Bilous2018} would apply to single sub-pulses rather than entire driftbands.

The paper is laid out as follows.
In \S\ref{sec:observations}, we describe our observational data, and how it was processed.
The adaptation of \citet{Bilous2018}'s arguments to develop a model that can be applied to single pulses is then described in \S\ref{sec:model} and applied to our uGMRT data set in \S\ref{sec:analysis}.
In \S\ref{sec:results}, we argue that the observed features of \psr{} are inconsistent with its assumed carousel being circular, and search for elliptical carousels that fit the model.
We also show how the model is sensitive to various assumptions that affect the symmetry of \psr{}'s profile evolution.
The discussion and conclusions are presented in \S\ref{sec:discussion}.

\section{Observations} \label{sec:observations}

\begin{figure}[t]
    \centering
    \includegraphics[width=1.0\linewidth]{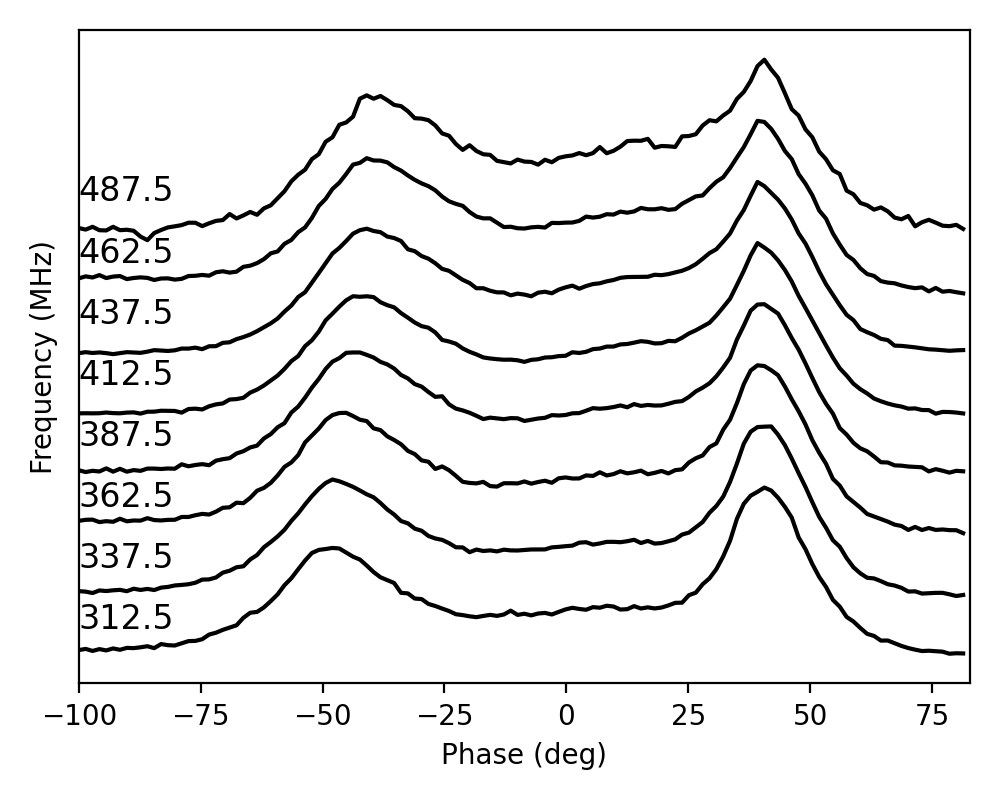}
    \caption{The \psr{} average profile produced from ${\sim}1.5$ hours of uGMRT data, split into eight subbands, and folded and dedispersed using the parameters obtained from the ATNF Pulsar Catalogue. The profile evolution is clearly visible, with the leading component approaching the trailing component at higher frequencies.}
    \label{fig:gmrt_pulsestack}
\end{figure}

The primary data set used in this work was taken using Band 3 ($300$-$500\,$MHz) of the upgraded Giant Metrewave Radio Telescope (uGMRT) located in Pune, India \citep{Reddy2017,Gupta2017}.
We used 13 antennas in the phased-array mode, recording total intensity data at a time resolution of $163.84\,\mu$s, and a frequency resolution of $48.8\,$kHz.

To supplement the component separation analysis at low frequencies (see \S\ref{sec:thorsett}), two observations of \psr{} were made with the Murchison Widefield Array \citep[MWA;][]{Tingay2013}.
Both used the Voltage Capture System \citep[VCS;][]{Tremblay2015} which recorded the raw voltages from 128 tile elements at a time resolution of $100\,\mu$s and a frequency resolution of $10\,$kHz, across a total bandwidth of $30.72\,$MHz.
These data were then beamformed towards \psr{}, using the software described in \citet{Ord2019}, and written out to PSRFITS files \citep{Hotan2004}.

Details of both the MWA and uGMRT observations are listed in Table \ref{tbl:observations}.

\begin{deluxetable}{l|ccc}
    \tablecaption{Observation metadata\label{tbl:observations}}
    \tablehead{
        \colhead{} &
        \colhead{MWA} &
        \colhead{MWA} &
        \colhead{uGMRT}
    }
    \startdata
        MJD & 57694 & 58613 & 57877 \\
        Bandwidth (MHz) & 30.72 & 30.72 & 200 \\
        No. of channels & 3072 & 3072 & 4096 \\
        Ctr. Freq. (MHz) & 184.96 & 215.68 & 400 \\
        Duration (s) & 1274 & 422 & 5273 \\
        No. of pulses & 2335 & 773 & 9660 \\
        Polarization & IQUV & IQUV & I
    \enddata
\end{deluxetable}

All the data sets were then dedispersed and folded using the standard pulsar software package, DSPSR \citep{Straten2011}.
The first round of DSPSR processing used the pulsar ephemeris published in the ATNF Pulsar Catalogue\footnote{\url{http://www.atnf.csiro.au/research/pulsar/psrcat}} \citep{Manchester2005}, and boosted the single pulse signal-to-noise (S/N) by averaging time bins to a final time resolution of $2.1\,$ms (equal to 1024 milliperiods).
The MWA data, for which the single pulse S/N was significantly smaller, was further time-averaged by another factor of 4 (i.e. 256 milliperiods).

As described below in \S\ref{sec:dm_determination}, the data were later reprocessed with identical parameters except for a different dispersion measure (DM).
After folding, the pulsestack data were written out using the software package PSRCHIVE \citep{Straten2012}.

\begin{figure*}[t]
    \centering
    \includegraphics[width=1\linewidth]{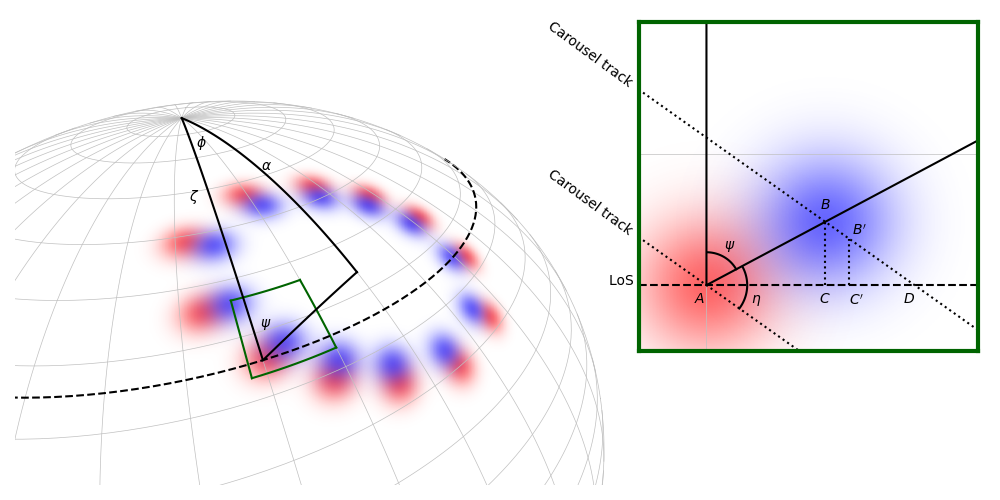}
    \caption{A schematic diagram illustrating the geometric model used in this work. The wire mesh sphere on the left represents the pulsar's sky, with the lines of ``longitude'' converging on the pulsar's rotation axis. The magnetic axis is at colatitude $\alpha$, while the LoS traverses the dashed circle at colatitude $\zeta$. An elliptical carousel of beamlets is drawn at an arbitrary reference frequency (red) and at a higher frequency (blue). Both are manifestations of only the outer carousel modelled in \citet{Bhattacharyya2009}, corresponding to the outer components of the profile. The inset shows the planar approximation (valid for small changes in frequency) of the geometry in the vicinity of the intersection of the carousel and the LoS. The meaning and interpretation of all other symbols are given in the main text.}
    \label{fig:model_geometry}
\end{figure*}

The data were then manually inspected for radio frequency interference (RFI).
The MWA data set required no RFI mitigation, but the uGMRT data showed multiple instances of both impulsive RFI affecting individual pulses, and persistent RFI in several channels.
Impulsive RFI was only flagged if it occurred in the pulse window and would affect the measurements of sub-pulse positions, resulting in 68 pulses being flagged.
After flagging, the uGMRT average profile had a S/N of ${\sim}1570$, equivalent to a single pulse S/N of ${\sim}16$.
The average profiles of this data set, split into eight subbands, are shown in Fig. \ref{fig:model_geometry}.

\section{Geometric Model} \label{sec:model}

The key idea underpinning our model, based on the carousel model, is that the longitude of a sub-pulse observed at a given frequency depends only on
\begin{enumerate}
    \item the location of the associated beamlet at that frequency, which by assumption must lie on a great circle that passes through the magnetic axis;
    \item the shape of the beamlet, which we assume here to be radially symmetric \citep[we do not consider other shapes, such as the fan beam model of][]{Dyks2017};
    \item the obliqueness of the LoS cut through the carousel track; and
    \item the motion of the beamlets due to the carousel's rotation.
\end{enumerate}

Fig. \ref{fig:model_geometry} illustrates the geometric model for a general elliptical carousel.
Suppose at frequency \vref{} the track intersects the LoS at point $A$ on the spherical surface representing the pulsar's sky (see Fig. \ref{fig:model_geometry}).
This point, along with the points representing the magnetic and rotation axes, forms a spherical triangle whose sides $\alpha$ and $\zeta$ meet at the rotation axis with angular separation $\phase$.

When a beamlet at frequency $\vref$ passes point $A$, a higher frequency counterpart of the same beamlet is at point $B$, following its own carousel track that intersects the LoS at point $D$.
If the carousel were stationary, then the center of the higher frequency sub-pulse would be observed at the phase corresponding to point $C$.
However, in the finite time it takes the LoS to move from $A$ towards $C$, the beamlet will have progressed some finite distance along its track, and will be observed at $C^\prime$.
In the diagram, $C^\prime$ is drawn on the right side of $C$, but it could also be on the left, depending on the direction of the carousel's rotation.

Relative to the reference frequency, let the observed phase shift of the average component be denoted $\Dphc$, and the observed shift of the sub-pulse be denoted $\Dphs$, which are therefore related to the geometry by
\begin{equation}
    \frac{\Dphs}{\Dphc} = \frac{\overline{AC^\prime}}{\overline{AD}}.
\end{equation}
For an infinitesimal change in frequency, this is also the ratio of the sub-pulse and component frequency evolution,
\begin{equation}
    \lim_{\Delta\nu \rightarrow 0} \frac{\Dphs}{\Dphc}
        = \frac{\text{d}\phase_s/\text{d}\nu}{\text{d}\phase_c/\text{d}\nu},
\end{equation}
evaluated at $\vref$.

\subsection{Stationary carousels}
\label{sec:stationary_carousels}

To derive an analytic expression for this ratio in terms of the local geometry, we first consider the case of a stationary carousel, when $C$ and $C^\prime$ coincide, and defer the generalization to a rotating carousel until the next subsection.
In the limit of a small change in frequency, the triangle $\triangle ABD$ becomes infinitesimally small, and can be approximated by a planar triangle, shown in the inset of Fig. \ref{fig:model_geometry}.
The ``local shape of the carousel'' can then be characterized by the angle $\eta$, defined in the Figure.
According to this definition, a perfectly circular carousel necessarily has $\eta = \pi/2$, while any other shape, such as an ellipse, will generally differ.

Applying the sine rule to $\triangle ACB$ yields $\sin\psi = \overline{AC} / \overline{AB}$, and applying the sine rule to $\triangle ABD$ gives
\begin{equation}
    \frac{\sin\angle ABD}{\overline{AD}}
        = \frac{\sin\angle ADB}{\overline{AB}}
        = \frac{\sin\psi \sin\angle ADB}{\overline{AC}}
    \label{eqn:trig1}
\end{equation}
Since $\angle ADB = \psi + \eta - \pi/2$, we can write $\sin\angle ADB = \cos(\psi + \eta)$.
Similarly, $\angle ABD = \pi - \eta$, and therefore $\sin\angle ABD = -\sin\eta$.
Therefore, Eq. \eqref{eqn:trig1} simplifies to
\begin{equation}
    \begin{aligned}
        \frac{\overline{AC}}{\overline{AD}}
            &= -\frac{\sin\psi \cos(\psi + \eta)}{\sin\eta} \\
            &= -\frac{\sin\psi}{\sin\eta} (-\sin\psi \sin\eta + \cos\psi \cos\eta) \\
            &= \sin^2\psi - \sin\psi\cos\psi\cot\eta.
    \end{aligned}
    \label{eqn:model_stationary}
\end{equation}

Eq. \eqref{eqn:model_stationary} depends on the pulsar's geometry via
\begin{equation}
    \tPs = \frac{\sAl\sin(\phase - \phase_0)}{\sZe\cAl - \cZe\sAl\cos(\phase - \phase_0)},
    \label{eqn:psi}
\end{equation}
where $\phase_0$ is the phase of the fiducial point, and so the problem of determining $\eta$ is only tractable if $\alpha$, $\zeta$, and $\phase_0$ are known, or can be constrained.
However, $\psi$ is also the angle typically associated with the PA (i.e. in the RVM), and if one is willing to accept this association, $\psi$ can be measured directly within the pulse window (given $\phase_0$) even if the RVM fit does not constrain $\alpha$ and $\zeta$ very well.
The second term on the right hand side of the last line of Eq. \eqref{eqn:model_stationary} encapsulates the deviation of the carousel from (local) circularity, which can be seen by noting that it vanishes when $|\eta| = 90^\circ$.

\subsection{Rotating carousels}

Eq. \eqref{eqn:model_stationary} relates the ratio of sub-pulse to average component evolution to the geometry of the pulsar and its carousel in the case that the carousel is not rotating.
The general case of a rotating carousel differs by a factor of $\overline{AC^\prime}/\overline{AC}$, which from Fig. \ref{fig:model_geometry} can be seen to represent the excess time it takes the LoS to ``catch up'' to the higher frequency manifestation of the same beamlet.
This factor depends on both the pulsar geometry and the carousel rotation, and in general can only be calculated if all of these quantities are known a priori.

On the other hand, the quantity $\overline{AC^\prime}/\overline{AC}$ is completely determined by the relative rates by which the sub-pulses and the visible point progress in phase, which can be inferred directly from the drifting behaviour.
Specifically, the time it takes the beamlet to move from $B$ to $B^\prime$ (and, equivalently, for its projection onto the LoS to move from $C$ to $C^\prime$) is equal to the phase difference associated with the line segment $\overline{CC^\prime} = \overline{AC^\prime} - \overline{AC}$, divided by the observed drift rate.
This must also be equal to the time it takes the pulsar to rotate through the phase associated with the line segment $\overline{AC^\prime}$.
All in all,
\begin{equation}
    \frac{\overline{AC^\prime} - \overline{AC}}{\delta}
        = \overline{AC^\prime} \frac{P}{2\pi},
\end{equation}
where $\delta$ is the drift rate expressed in units of rad/s.
Rearranging,
\begin{equation}
    \frac{\overline{AC^\prime}}{\overline{AC}} = \frac{1}{1 - \frac{P}{2\pi} \delta}.
\end{equation}

The use of the term ``drift rate'' for the symbol $\delta$ used here differs slightly from the usual meaning of the term in the context of sub-pulse drifting.
If a sub-pulse is observed to shift in phase by $\DphDR$ from one pulse to the next, it is said to have a drift rate of ``$\DphDR$ per pulse''.
However, this definition is only an approximation to the rate at which the sub-pulse progresses through phase (which is what we mean here by $\delta$), because it fails to take into account the fact that the sub-pulse traversed $\DphDR$, not in time $P$, but in time $P + \DphDR P/(2\pi)$.
Thus, the drift rate is related to the observed change in phase by
\begin{equation}
    \delta = \frac{\DphDR}{P\left(1 + \frac{\DphDR}{2\pi}\right)},
\end{equation}
or, when the quantity $\DphDR$ is approximated by $2\pi P_2/P_3$ (where $P_2$ and $P_3$ are both expressed in equivalent units of time),
\begin{equation}
    \delta = \frac{2\pi}{P} \frac{P_2}{P_3 + P_2},
\end{equation}
and therefore
\begin{equation}
    \frac{\overline{AC^\prime}}{\overline{AC}}
        = 1 + \frac{P_2}{P_3}
\end{equation}
If the drift rate is aliased with aliasing order $k$, then the sub-pulse traverses an extra phase $2\pi k P_2/P$ in a correspondingly longer or shorter time (depending on the sign of $k$), yielding the final result
\begin{equation}
    \frac{\overline{AC^\prime}}{\overline{AC}}
        = 1 + \frac{P_2}{P_3} + \frac{kP_2}{P}.
\end{equation}

Incorporating this correction into the model presented in \S\ref{sec:stationary_carousels}, we arrive at a complete expression for the ratio of the evolution rates of sub-pulses to average components that takes into account the pulsar's geometry, the local geometry of the carousel, and the carousel rotation:
\begin{equation}
    \frac{\text{d}\phase_s/\text{d}\nu}{\text{d}\phase_c/\text{d}\nu}
        = (\sin^2\psi - \sin\psi\cos\psi\cot\eta) \left( 1 + \frac{P_2}{P_3} + \frac{kP_2}{P} \right).
    \label{eqn:model}
\end{equation}

\section{Analysis} \label{sec:analysis}

\subsection{Dispersion measure determination}
\label{sec:dm_determination}

Using the model outlined in the previous section to test the circularity of \psr{}'s carousel involves the measurement of the frequency evolution of both the average profile components and the individual sub-pulses.
However, pulsar signals are dispersed, primarily by the interstellar medium, causing a delay inversely proportional to the square of the frequency.
The proportionality constant is commonly decomposed into a factor that describes the physics of dedispersion in cold plasmas ($\mathcal{D} \approx 4.15 \times 10^3\,$MHz$^2\,$pc$^{-1}\,$cm$^3\,$s), and a factor that describes the amount of dispersing material along the line of sight, commonly called the \textit{dispersion measure}, or DM, given in units of \DMunits.

A correct characterisation of the frequency evolution therefore requires that the data have been dedispersed by the correct amount, as any residual dispersion will change the apparent rate at which components (or sub-pulses) shift in phase with frequency.
In particular, a non-zero residual dispersion, $\DDM$, will add an extra slope such that the measured quantities (indicated by tildes) will differ from the true quantities according to
\begin{equation}
    \Deriv{\tilde{\phase}_{s/c}}{\nu} = \Deriv{\phase_{s/c}}{\nu} + \Deriv{\phDM}{\nu},
    \label{eqn:DMcorrection}
\end{equation}
valid for both sub-pulses (``$s$'' subscript) and average components (``$c$'' subscript), where
\begin{equation*}
    \deriv{\phDM}{\nu}
        = \frac{2\pi}{P}\deriv{t}{\nu}
        = -\frac{4\pi}{P}\frac{\mathcal{D}\times\DDM}{\nu^3}.
\end{equation*}

In the presence of profile evolution, measuring the DM is fundamentally ambiguous, unless the nature of the evolution is known a priori.
Standard S/N-maximizing algorithms for determining the DM are not robust to asymmetric profiles in which one component dominates \citep{Ahuja2007}.
In such cases, the DM will be biased towards the value that lines up the brightest component across the observed frequency range.

Although frequency evolution is less prominent at higher frequencies where DM measurements are often made \citep{Thorsett1992}, it is likely that DM measurements of \psr{}, whose profile becomes increasingly asymmetric below ${\sim}1\,$GHz, are biased.
Fig. \ref{fig:gmrt_pulsestack} was made using the catalog DM of $113.4\,$\DMunits, originally obtained by \citet{Arzoumanian1994} using observations at $0.4$, $0.8$, and $1.33\,$GHz.
The fact that the trailing component is aligned, while the leading component shows significant evolution, implies that the DM is likely overestimated in this case.

Our own measurement of the DM, using the pdmp tool in PSRCHIVE \citep{Straten2012}, yielded $113.185 \pm 0.031$\DMunits{} in our uGMRT data set, which is marginally consistent with the catalog value whose stated uncertainty is $0.2\,$\DMunits.
Although lower, this DM also produces a profile stack (not shown here) in which the evolution of the leading component is more pronounced than the trailing component.

Neglecting other effects (such as aberration and retardation, hereafter AR), the two components arising from a LoS cut through a circular emission cone are expected to exhibit symmetric frequency evolution.
Since we are investigating the possibility that \psr{}'s carousel is \emph{not} circular, we start with the ansatz that the average profile evolution \emph{is} symmetric, and test whether this is consistent with the observed sub-pulse evolution.
Accordingly, after fitting the profile components (described in the following section), we found the DM that produced the most symmetric average profile evolution by fitting a slope to the midpoints between the component positions in each subband, and subtracted the DM that would generate this slope.
The resulting symmetry-maximising DM, which was found to be $\DMsym = 113.098\,$\DMunits, was used to re-process all of the data for the initial analysis.


\subsection{Component and sub-pulse fitting}

The proposed emission beam of \citet{Bhattacharyya2009} is a pair of nested carousels, with the larger carousel corresponding to the prominent outer profile components and the smaller carousel, which the LoS cuts through tangentially, corresponding to the inner component, or bridge emission.
Although our sub-pulse model is equally valid for both carousels, measuring the frequency behavior of the bridge emission is difficult, due to the difficulty of isolating the bridge emission from the outer components unambiguously.
Consequently, in this study we restrict ourselves exclusively to the outer components, and hence the larger carousel, whose profile evolution is easier to measure unambiguously.

The superb S/N allowed to divide the total $200\,$MHz ($= 4096 \times 48.8\,$kHz channels) into a large number of subbands, while still retaining a large enough S/N in each subband to measure the component peak positions very precisely.
We found that 64 subbands ($3.125\,$MHz each) provided an adequate compromise of S/N and frequency resolution.
A small number of these subbands were flagged either because of contaminating RFI or because of the degradation of the signal due to the bandpass of the Band 3 receiver (at the top and bottom of the $200\,$MHz range).

\begin{figure}[t]
    \centering
    \includegraphics[width=\linewidth]{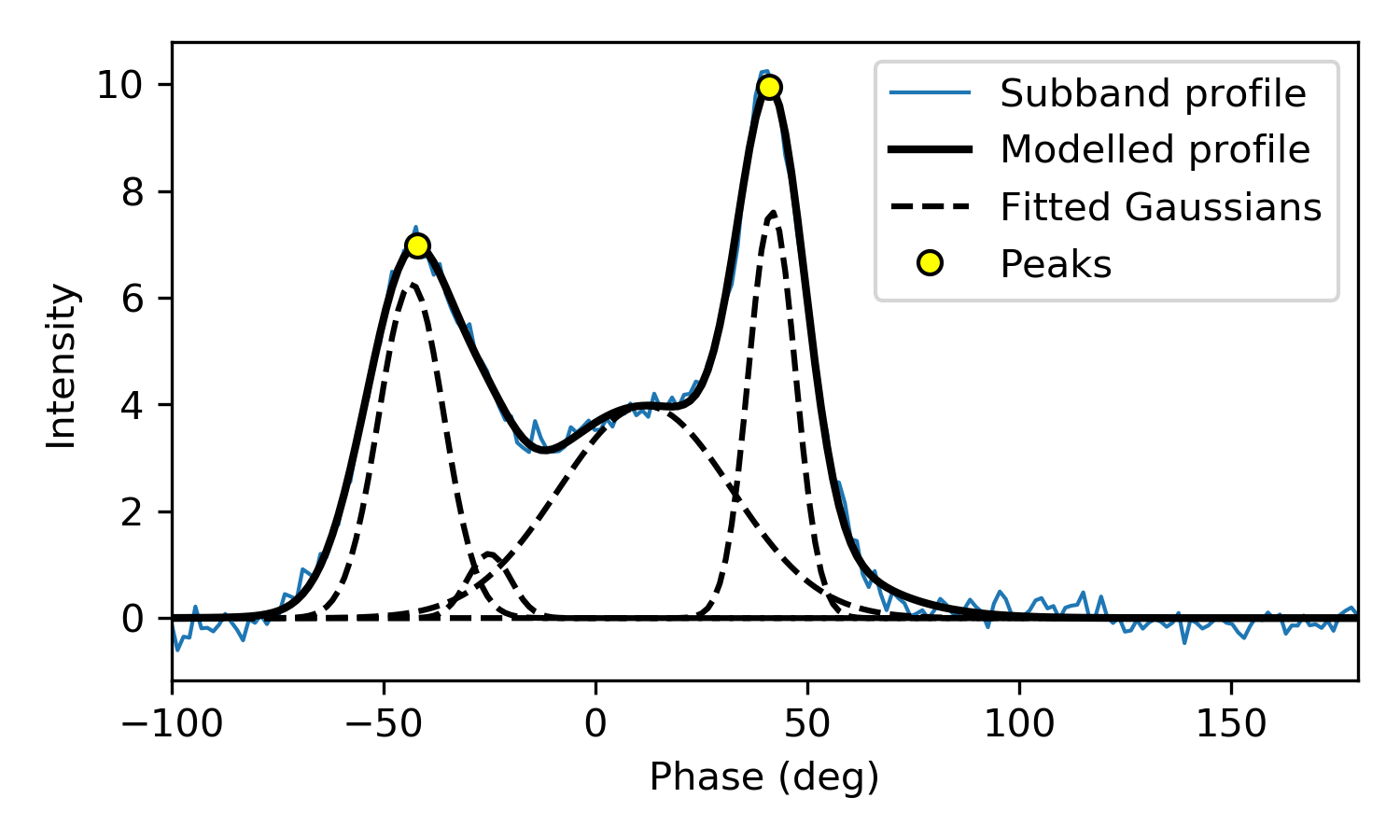}\\
    \includegraphics[width=\linewidth]{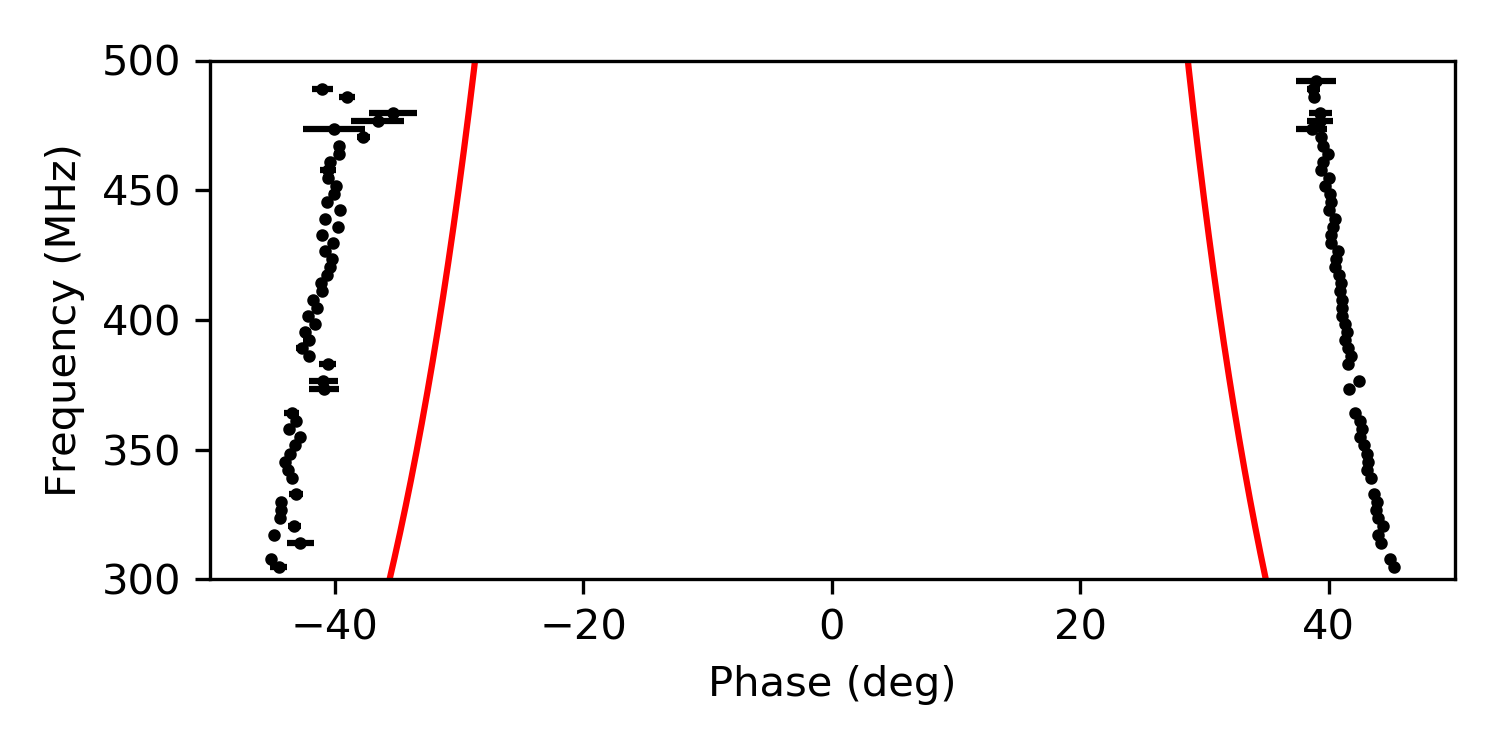}
    \caption{\textbf{Top}: A demonstration of the fitting procedure applied (described in the text) to one of the subband profiles. The original data are shown along with the fitted Gaussian components, the resulting modelled profile, and the peaks of the outer components. \textbf{Bottom:} The fitted peak positions (black points) for all subbands from $300$ to $500\,$MHz for which fits could be obtained. The red lines are power law fits to the evolution according to Eq. \eqref{eqn:powerlaw}, displaced by $\pm 10^\circ$ so as not to obscure the peak positions.}
    \label{fig:example_profile_fitting}
\end{figure}

Each subband profile was then fitted with four Gaussians simultaneously.
We found that at least four components were needed to faithfully reconstruct the entire profile across the whole pulse window, for all the subbands.
After the Gaussian fitting, model profiles were constructed by summing the fitted components together.
The local maxima of the model profiles were then found (using a cubic spline for interpolation), only retaining the local maxima that corresponded to the outer components, which was checked manually.
The top panel of Fig. \ref{fig:example_profile_fitting} illustrates the results of this peak-finding procedure for one of the subbands.

The uncertainty of the peak positions is propagated from the covariance matrix of the Gaussian fits.
Since the peak positions are defined as the phase at which the first derivative vanishes, we first calculate the error on the first derivative of the model as
\begin{equation}
    \sigma_y = \sqrt{{\bf J}_y{\bm \Sigma}{\bf J}_y^\text{T}},
    \label{eqn:error_prop}
\end{equation}
where $y = \text{d}I/\text{d}\phase$ represents the derivative of the profile intensity, $I$, as a function of phase, ${\bm \Sigma}$ is the covariance matrix of the model fit (with 12 parameters, 3 for each Gaussian component), and ${\bf J}_y = [\text{d}y/\text{d}a_1, \text{d}y/\text{d}a_2, \dots]$ is the Jacobian matrix with respect to each of the model fit parameters, $a_1, a_2, \dots, a_{12}$.
We then assume that the first derivative of the model is sufficiently smooth in the vicinity of the peaks that the errors in the peak position are well approximated by dividing $\sigma_y$ by the gradient of $y$, or the second derivative of the model,
\begin{equation}
    \sigma_\phase \approx \sigma_y \left(\dd{I}{\phase}\right)^{-1},
\end{equation}
evaluated at each peak position.
The final set of fitted peak positions for all subbands with sufficiently high S/N is shown, with their errors, in the bottom panel of Fig. \ref{fig:example_profile_fitting}.

The positions of the individual sub-pulses were determined in a similar way.
However, to maintain a sufficiently high S/N, the single pulses were divided into $8 \times 25\,$MHz subbands (instead of the 64 subbands used for the average profile components).
Four Gaussians were still used for fitting each subband, except that fits were rejected if no peak occured above three standard deviations of the residual noise.

\subsection{Component and sub-pulse evolution}

For the purposes of the sub-pulse model, we are interested in the comparison between the instantaneous evolution of the average profile components and that of the sub-pulses at one particular reference frequency, which we choose to be the center of our band, $\vref = 400\,$MHz.
As can be seen in Fig. \ref{fig:example_profile_fitting}, the observing bandwidth is large enough to see a slightly non-linear trend of phase with frequency, especially in the brighter trailing component where the peak fitting uncertainties are smaller.
Thus, obtaining the instantaneous slope at $400\,$MHz requires something more sophisticated than simply fitting a straight line through the peak positions.

On the other hand, the lower S/N of the sub-pulses meant that often peak positions were only reliably measurable in a relatively small number of subbands, whose evolution would therefore be in danger of being overfit by models with too many free parameters.
However, it is preferable to use only a single model for both the average profile components and the sub-pulses for consistency.
Therefore, to retain both the small number of parameters needed for the sub-pulses, and still be able to fit the curvature, we chose the two-parameter power law,
\begin{equation}
    \phase = A \nu^B,
    \label{eqn:powerlaw}
\end{equation}
where $A$ and $B$ are the free parameters to be fitted.

The precise fitted values of $A$ and $B$ are not physically important, but serve to provide a robust measurement of the instantaneous evolution at our chosen reference frequency.
The least squares fits to both average profile components are shown in Fig. \ref{fig:example_profile_fitting}.
The fitted power law can then be differentiated to obtain an instantaneous evolution measurement at any desired reference frequency,
\begin{equation}
    \left.\deriv{\phase}{\nu}\right|_{\nu = \vref} = AB\vref^{B-1},
\end{equation}
with uncertainty calculated from standard assumed normal statistics, analogous to Eq. \eqref{eqn:error_prop}.

Because of the lower S/N, obtaining similar power law fits to the sub-pulse evolution required more care.
Occasionally, two or more narrow sub-pulses occured within the phase range associated with a single average component, and the Gaussian fitting procedure selected different sub-pulses in different subbands, which would yield an inaccurate measurement of the evolution.
We therefore rejected any subband whose sub-pulse peak position was more than two standard deviations away from the average phase of the sub-pulse positions within that pulse across the subbands.
After this round of pruning subbands, we then rejected any sub-pulse that did not have at least four subbands with well-determined peak positions.
This procedure yielded 1176 suitable sub-pulses associated with the leading average component, and 1566 sub-pulses associated with the trailing average component.
These were individually inspected for further cases of misclassified sub-pulse fits (for example, due to unflagged RFI), resulting in a final yield of 1131 sub-pulses in the leading component, and 1543 in the trailing component, which were then fitted with power laws to model their evolution across frequency.

\section{Results} \label{sec:results}

\begin{figure}[t]
    \centering
    \includegraphics[width=\linewidth]{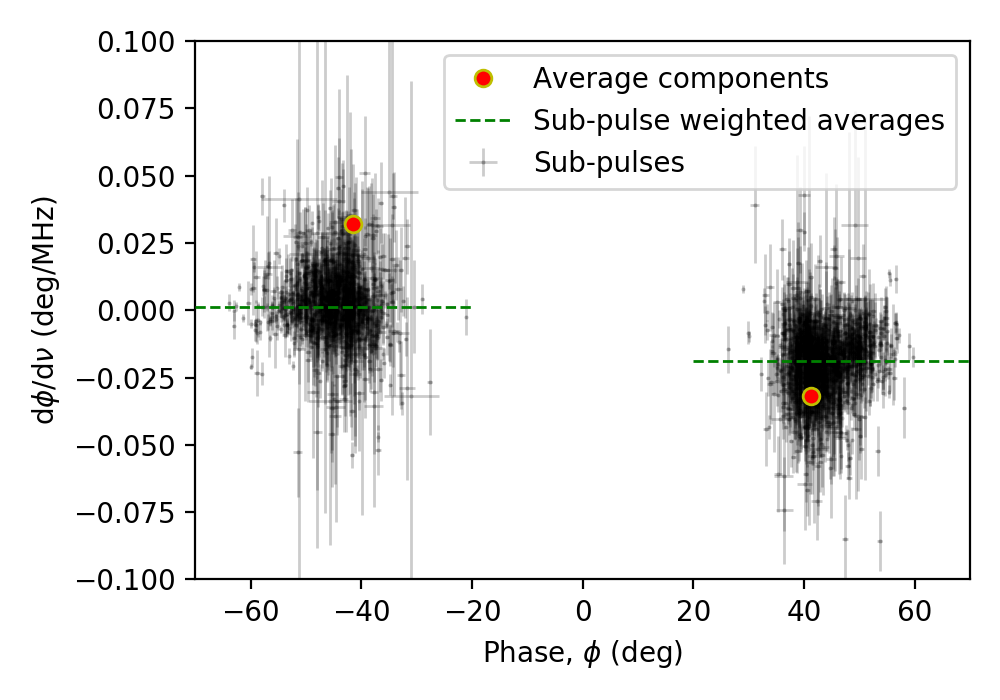}
    \caption{The phases and evolutionary behaviour of both the average profile components and individual sub-pulses measured at 400\,MHz, using $\DM = 11.098\,$\DMunits. The error bars of the average profile components are too small to be seen at this scale and are therefore not shown. The weighted averages of the sub-pulses' evolution ($\text{d}\phase_s/\text{d}\nu$) are indicated by the horizontal green dashed lines.}
    \label{fig:PS_both}
\end{figure}

\psr{} was measured by \citet{Bhattacharyya2007} to have $P_2 \approx 28^\circ$ for the outer components and $P_3 \approx 18.3\,P$ near our chosen reference frequency of $\vref = 400\,$MHz.
For an unaliased carousel rotation rate ($k = 0$), this introduces a ${\sim}0.4\%$ correction to the model, as per the seconded bracketed expression in Eq. \eqref{eqn:model}, while for first order aliasing ($|k| = 1$), this amounts to an ${\sim}8\%$ correction.
In the following, we use $k = 1$ to be consistent with \citet{Bhattacharyya2007}, but it should be noted that the effect on the final measurement of $\eta$ is marginal, and so we make no further attempt to make this calculation more precise by measuring $P_2$ and $P_3$ in our own data set.

The phases and frequency evolution ($\text{d}\phase/\text{d}\nu$) of both the average profile components and the individual sub-pulses, determined by the analysis method described above, are plotted in Fig. \ref{fig:PS_both}.
Clearly, there is a significant difference between the evolutionary behaviour of the average profile components and the sub-pulses, with the average profile components showing the more pronounced evolution.
Moreover, the average sub-pulse behaviour is different for the two components, with the trailing component showing the more pronounced sub-pulse evolution on average.
This observation alone implies that the assumed carousel cannot be symmetrical, and hence cannot be circular.

This conclusion is borne out by the values of $\eta$ calculated from Eq. \eqref{eqn:model}, and using the $\psi$ values obtained from a known viewing geometry via Eq. \eqref{eqn:psi}.
\citet{Bhattacharyya2009} found two geometries arguably consistent with the PA sweep, but only one of them ($\alpha \approx 175.^\circ4$, $\beta \approx -6.^\circ9$) is consistent with the observed PA of the outer components.
The results of these calculations are presented in Table \ref{tbl:main_results}, which show that $\eta$ differs markedly from the $\pm 90^\circ$ expected for a circular carousel.
We must therefore entertain the possibility of non-circular carousel shapes.

\begin{deluxetable}{cCCCCC}
    \tablecaption{The carousel geometry implied by the sub-pulse model under initial assumptions.\label{tbl:main_results}}
    \tablehead{
        \colhead{Component} &
        \dcolhead{\psi} &
        \dcolhead{\deriv{\phase_s}{\nu}} &
        \dcolhead{\deriv{\phase_c}{\nu}} &
        \dcolhead{\eta} \\
        \colhead{} &
        \colhead{($^\circ$)} &
        \colhead{($^\circ$/MHz)} &
        \colhead{($^\circ$/MHz)} &
        \colhead{($^\circ$)}
    }
    \startdata
        Leading & -21 & 0.00117 & 0.032 \pm 0.001 & -75 \\
        Trailing & 21 & -0.0187 & -0.032 \pm 0.001 & -33
    \enddata
\end{deluxetable}

\subsection{Elliptical carousels}

Another carousel shape that has been considered in the literature is an ellipse, e.g. by \citet{Wright2017a} to account for the phenomenon of bi-drifting.
In justifiying the choice of an ellipse, they point out that there is no a priori reason to prefer any particular shape, not even a circular one, and ellipses are the natural choice in cases where circles prove inadequate, because they represent a minimal step-up in complexity.
We thus consider whether an ellipse can account for the observed asymmetry of component vs sub-pulse evolution under our stated assumptions.

Note that the $\eta$ values given in Table \ref{tbl:main_results} have the same (negative) sign.
Since $\eta$ is only uniquely defined within the range $-90^\circ \le \eta \le 90^\circ$, these values indicate that the orientation of the carousel track at the two intersection points is rotated by an acute angle in the \emph{same} direction, relative to the orientation implied by a circular carousel.
Taken at face value, this implies a very peculiar carousel shape, since even ellipses will generally have the property that the orientation at two points of intersection will differ from circularity in opposite senses, unless the ellipse is eccentric with its major (or minor) axis rotated significantly relative to the fiducial plane.

However, even if a rotated, elliptical carousel can be found which agrees with the above estimates of $\eta$, such a geometry implies that the fiducial point is \emph{not} located midway between the two average components, but off to one side or the other, contrary to what was assumed in the above calculation.
Furthermore, an elliptical, rotated ellipse would intersect the LoS in an asymmetric way, and would exhibit a generally asymmetric evolution, which contradicts the assumption underlying our choice of DM.
Therefore, the search for a viable elliptical geometry requires us to revisit our choice of DM.

\subsubsection{Revisiting the DM}

Whereas our original choice of DM was fixed by the ansatz that the carousel was circular (and the profile components therefore evolving in a symmetric way), we now seek a DM that is consistent with the expected frequency evolution of an elliptical carousel.
However, ellipses have five degrees of freedom (in both planar and spherical geometries), only four of which are constrained by our measurements and assumptions: (1) the longitudinal separation of the components, (2) the tangent of the ellipse at the leading LoS intersection point, (3) the equivalent tangent on the trailing side, and (4) the magnetic inclination, $\alpha$.
Thus, for every choice of DM, there will be a whole family of viable ellipses, from which a single solution can only be selected if some extra constraining assumption is made.

The last constraint is tantamount to fixing the longitude of the fiducial point, $\phi_0$.
Usually, $\phi_0$ is estimated either using the profile's symmetry (as was done in our initial analysis), or by using the RVM fit to the polarization curve.
However, in the present context, symmetry arguments do not apply, since for every choice of $\phi_0$ a possible ellipse can be found.
Furthermore, our observation was only total intensity, and so we cannot use it to fit the RVM to the PA curve, or to interrogate the uncertainties of previously reported fits.
Indeed, relying too heavily on the PA curve to determing $\phi_0$ is problematic in any case, since \psr{}'s PA curve is known not to fit any RVM curve well, as shown by \citet{Bhattacharyya2009}.

Nevertheless, we can restrict the set of possible solutions by testing which ranges of DM and $\phi_0$ yield reasonable geometries.
There is considerable freedom here, especially given the pulsar's complex PA curve, which is arguably consistent with a broader range of viewing geometries, $(\alpha, \beta)$, than the two solutions published by \citet{Bhattacharyya2009}.
Given this freedom, we will favour solutions that predict ellipses with minimal eccentricity.
This immediately implies that small adjustments of $\phi_0$ (from the symmetric center when $\DMsym = 113.098\,$\DMunits) are preferred, since larger displacements increases the asymmetry of the average profile components, requiring the ellipse to be increasingly eccentric and rotated with respect to the fiducial plane.
Moreover, since changing $\phi_0$ also requires that $\alpha$ and $\beta$ be adjusted in order to retain reasonable RVM fits to the PA curve, we will seek solutions with fiducial points not more than a few degrees from the fiducial point used in the RVM fits published in \citet{Bhattacharyya2009}.

We therefore searched for possible solutions within the DM range $\DMsym \pm 0.25\,$\DMunits and restricting $\phi_0$ to within $3^\circ$ of the symmetric center (when dedispersed to $\DM = \DMsym$, as shown in Fig. \ref{fig:example_profile_fitting}).
For each DM, the correction to the measured slopes were calculated according to Eq. \eqref{eqn:DMcorrection}, while for each $\phi_0$, the values of $\psi$ for both the leading ($\psi_L$) and trailing ($\psi_T$) components needed to be adjusted, according to the different geometry implied by the shift of the fiducial plane.
If polarisation data are available, this could be achieved by refitting the RVM to the PA curve for each $\phi_0$.
Since our data were only total intensity, we estimated the new $\psi_L$ and $\psi_T$ by subtracting $\psi_0$ from the values listed in Table \ref{tbl:main_results}, where $\psi_0$ is the PA predicted by the original geometry at the (new) fiducial point.
This method gives results that only approximate the adjusted geometries, but it still ensures that the adjusted values of $\psi_L$ and $\psi_T$ are consistent with the PA curve itself.

\begin{figure*}[t]
    \centering
    \includegraphics[width=1\linewidth]{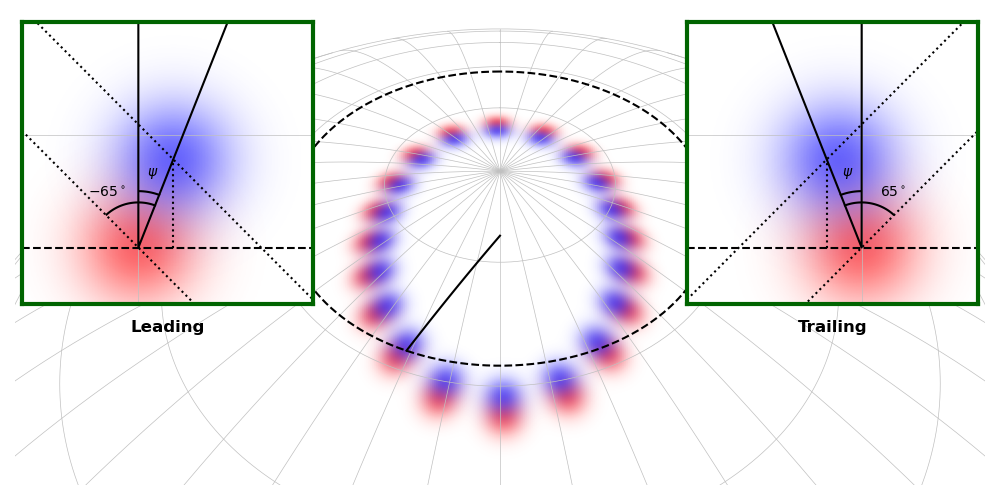}
    \caption{A representative solution of the carousel geometry of \psr{} in the style of Fig. \ref{fig:model_geometry}, again with the two colors representing different frequency manifestations of a single carousel. A single solid line is drawn from the (dashed) LoS track to the magnetic axis, and the angular distance from the magnetic axis to the nearest pole is $180^\circ - \alpha = 4.^\circ6$. Insets showing LoS cuts through the elliptical carousel on both leading and trailing sides of the profile. The ellipse is chosen to approximate $\eta_L \approx -65^\circ$ and $\eta_R \approx 65^\circ$, where the sign is indicative of the fact that $\eta$ is always defined in a clockwise sense from the bearing which points towards the magnetic axis. The shift of the fiducial point in this geometry is too small ($\lesssim 1^\circ$) to be easily seen in this image.}
    \label{fig:model_geometry_results}
\end{figure*}

Thus, for each choice of $\DM$ and $\phi_0$, the angle $\eta$ was recalculated for each component ($\eta_L$ for leading, $\eta_T$ for trailing).
As discussed earlier, minimally eccentric (and minimally rotated) ellipses are expected to have $\eta_L$ and $\eta_T$ be oppositely signed, and we found that such solutions only exist for DMs between $113.13$ and $113.3\,$\DMunits.
Among these solutions, the sign of $\eta_L$ is negative, implying an ellipse whose major axis is inclined closer to the fiducial plane than the minor axis (see Fig. \ref{fig:model_geometry_results} for an illustration of a representative solution).
Solutions with $|\eta_L| \approx |\eta_T|$ (expected when the LoS approaches a tangent to the ellipse) occur in the DM range $113.23$ to $113.27\,$\DMunits, with $|\eta|$ ranging from ${\sim}60^\circ$ to ${\sim}65^\circ$ for both components.
Even these so-called ``minimally eccentric'' ellipses show significant eccentricity, and the best fitting solution may lie slightly outside the ranges given above.

\subsection{Average profile component separation}
\label{sec:thorsett}

The model used above to derive possible carousel shapes uses the ratio of the average profile component evolution to the sub-pulse evolution, and is based on the assumption that this ratio remains constant regardless of how the beamlets' instantaneous positions change with frequency.
However, thanks to the broad frequency coverage and the high signal-to-noise, our data set can also be used to probe the evolution itself from $300$ to $500\,$MHz.
Here, we measure the phase separation between the two outer components, and combine it with similar published measurements at higher and lower frequencies to obtain a global picture of \psr{}'s evolution.
Unlike in the model above, the component separation does not depend on getting the DM correct, as long as the pulse morphology is not strongly affected by DM smear, which is the case here.

\begin{figure}[th]
    \centering
    \includegraphics[width=\linewidth]{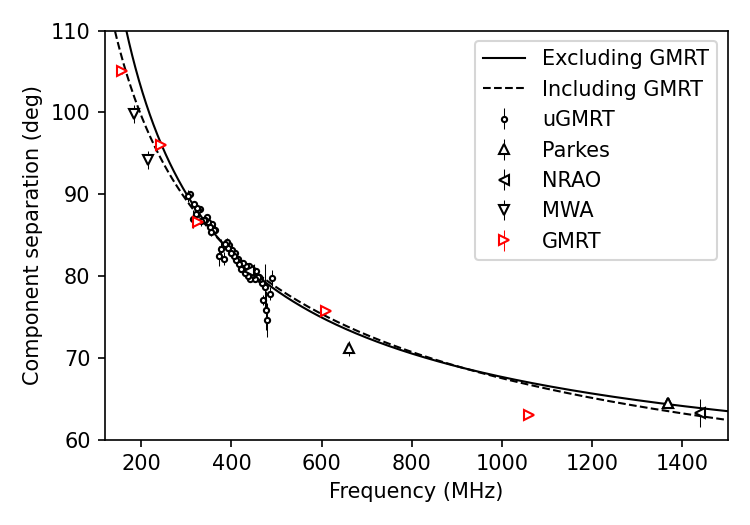}
    \caption{Measurements of the average profile component separations from our own observations (uGMRT and MWA), the EPN database from NRAO and Parkes, and the measurements published in \citet{Bhattacharyya2009} (GMRT, red). Both fitted lines use the empirical model advanced by \citet{Thorsett1992}, with the solid line including only measurements made using the method described in this paper (black points), and the dashed line including all the points.}
    \label{fig:thorsett}
\end{figure}

Fig. \ref{fig:thorsett} shows the average profile component separation measured at frequencies between ${\sim}150$ and ${\sim}1440\,$MHz.
The uGMRT and MWA points are from our own observations, while the NRAO \citep{Qiao1995} and Parkes \citep{Arzoumanian1994,Johnston2017} measurements were made using data downloaded from the European Pulsar Network database\footnote{\url{http://www.epta.eu.org/epndb/}}.
In all these cases, the average profile was fit with three Gaussians, unless the fit was visibly improved by adding an extra Gaussian component, in which case four were used.
The peak finding method and error analysis were identical to the method described in Section \S\ref{sec:analysis}.
In addition to these points, we have included the component separation measurements reported in \citet{Bhattacharyya2009}, displayed in red.
These measurements have been treated separately because the method used to measure the separation is not necessarily the same.

The average profile component separations were fit with the empirical model put forward by \citet{Thorsett1992}:
\begin{equation}
    \Delta\phase = A\nu^\alpha + \Delta\phase_\text{min},
\end{equation}
where $A$, $\alpha$ and $\Delta\phase_\text{min}$ are considered free parameters.
We performed two fits, one including only the component separations measured using our method (i.e. excluding the measurements of \citealt{Bhattacharyya2009}),
\begin{equation}
    \Delta\phase = (17 \pm 5)^\circ \times \left( \frac{\nu}{\text{GHz}} \right)^{-0.70 \pm 0.06} + (51 \pm 2)^\circ,
    \label{eqn:thorsett_fit_detail1}
\end{equation}
and one including all measurements,
\begin{equation}
    \Delta\phase = (31 \pm 5)^\circ \times \left( \frac{\nu}{\text{GHz}} \right)^{-0.44 \pm 0.04} + (36 \pm 4)^\circ.
    \label{eqn:thorsett_fit_detail2}
\end{equation}

Although the two sets of fitted parameters are not consistent with each other, they both produce models that describe the data reasonably well within the given frequency range, as is apparent in Fig. \ref{fig:thorsett}.
However, both fits are clearly dominated by the uGMRT points and the high-frequency points, with the other points generally falling below the fitted lines.
The tendency for the other points to be underestimated (relative to the models) can be partially understood by considering the effect of the profiles' finite bandwidths.
If the evolution is significant across the observing band, then the relative location of the components of the average profile are affected by their relative spectral indices.

In the case of \psr{}, the leading component has a shallower spectral index than the trailing component.
Therefore, the band-averaged leading component is weighted more heavily by its higher frequency part relative to the trailing component, and, since the leading component moves towards the trailing component at high frequencies, the net result is that the component width will be underestimated.
Because the GMRT components were measured in small subbands (relative to the evolution), it is likely that the effect just described is neglible for the GMRT data.
The effect will also be less significant for the highest frequency observations as well, since larger bandwidths would be needed before the effects of evolution become noticeable.
The low frequency MWA and GMRT measurements are likely also to be affected by scattering, which starts affecting the profile shape noticeably at frequencies $\lesssim 250\,$MHz, although it is not clear exactly how this will bias the component separation measurement.

In short, although the fits are clearly dominated by the uGMRT and high frequency Parkes measurements, there are grounds for considering these measurements to be the least underestimated, and therefore the most accurate.

\section{Discussion and Conclusions} \label{sec:discussion}

We have extended the model of \citet{Bilous2018} to explore the relationship between the frequency evolution of individual sub-pulses and the average profile components, and applied this model to uGMRT observations of \psr{} in the frequency range $300$-$500\,$MHz.
Although the analysis relies on the pulsar's profile evolution, which is commonly thought to arise from a RFM, the results are independent of any one particular RFM, as it is only the \textit{relative} evolution of the sub-pulses and the average components (specifically, the fractional difference of their evolution rates) that has any bearing in the model.
This relative quantity will not change if the evolution of the sub-pulses and the average components are governed by the same underlying principle, such as RFM.

The main result presented in this paper is that, if the observed drifting behaviour is indeed governed by a carousel of discrete, radially symmetric beamlets, then the carousel track cannot be circular.
This result depends on a number of assumptions, explicitly laid out at the beginning of \S\ref{sec:model}, any of which may turn out to be false.
Indeed, even the implicit assumption that the pulsar beam is carousel-like may be challenged, using either general arguments along the lines of \citet{Edwards2006}, or by suggesting alternative models \citep[e.g.][]{Clemens2004,Gogoberidze2005,Dyks2017,Basu2020}.
Furthermore, among the carousel-like models, multiple non-circular geometries are considered \citep{Biggs1992a,Narayan1983,Wright2017a}.

Similarly, the assumption that the beamlets are radially symmetric may prove too simplistic.
Often, in pulsar studies of this nature \citep[e.g.][]{Maan2019}, it is the \emph{sparks on the surface} that are assumed to be radially symmetric.
Under this assumption, the far-field beamlets that are ultimately observed are not necessarily radially symmetric, but have shapes that are governed by the details such as the size of the sparks, the size of the carousel, the distribution of particle energies at the observed emission heights, and whether AR effects therefore become important.
The combination of these effects may produce, e.g., beamlets that are elongated along the direction pointing radially outward from the magnetic axis, possibly with some curvature (if a higher multipole of the global magnetic field is strong enough at the emission heights).
The beamlet carousel may also exhibit a measurable phase lag at different frequencies, which would counter our assumption that the beamlets move radially inward toward the magnetic axis at higher frequencies.

Within this broad landscape of possible beam geometries, it is worth pausing to consider the usefulness of the analysis presented above.
Our analysis serves as a kind of blueprint for how specific predictions about the relationship between individual sub-pulse behaviour and average component behavior can be made.
We have taken a relatively simple beam geometry, and allowed it to vary in one particular way (the eccentricity of the beam carousel).
However, a similar approach could be taken to test individual aspects of a more comprehensive model that uses the underlying spark configuration as its starting point, by first working out how the beam geometry depends on those aspects.
The usefulness of such predictions as a litmus test for the viability of the carousel model itself is somewhat lessened by the sheer number of free parameters available, even within specific emission mechanisms such as curvature radiation.
However, even in our conceptually simpler approach, which obscures the details of the emission mechanism in favour of an assumed beam geometry, the freedom one has to freely ``extend'' the model to include elliptical carousels \emph{also} increases the number of free parameters in the model, making it more likely to be able to find a solution to fit any observed behavior.

Thus, although we have presented a possible geometry (or rather, a range of geometries) that may explain the asymmetry of sub-pulse and average component evolutionary behaviours for \psr{}, we emphasize that it is not the solution itself that is of primary importance, but rather the demonstration of how single pulse measurements can be exploited to provide extra constraints on a given model of the pulsar's beam (in this case, that the carousel cannot be circular).
Performing similar analyses in the context of other beam geometries may prove equally illuminating, potentially ruling out various families of beam geometries.

Our model, based on the carousel model, inherits its shortcomings as well.
For example, although our model deals with certain asymmetric features of \psr{}'s emission, it does not address what is arguably the most obvious asymmetry of them all: the relative brightness of the two main (outer) components, and their different spectral indices.
Profile asymmetry is a general phenomenon observed in most pulsars, whose explanation still has no widely accepted consensus.
In this respect, our model does no better or worse than the carousel model itself, which has no natural way to explain such asymmetries.

Apart from these deficiencies in the model itself, our analysis was hampered by a few practical difficulties.
The model was only applied to the outer profile components because the sub-pulses in the inner components, which \citet{Bhattacharyya2009} associated with a secondary carousel, did not have sufficiently high S/N to measure their evolution across the frequency range.
If such measurements could be made, e.g. with PTUSE\footnote{Pulsar Timing User Supplied Equipment} \citep{Bailes2020,Song2023}, the value of $\eta$ could be tracked across the pulse window, providing much stronger constraints on possible carousel geometries.

Another difficulty is the twin ambiguity of the DM (given the asymmetry of the profile) and of the fiducial point ($\phase_0$), related to \psr{}'s complex PA curve.
As has been shown, $\eta$ is sensitive to small changes to both the DM and $\phase_0$, making it difficult to identify a single ``correct'' carousel geometry.
However, despite this, our model appears to favor DM values slightly smaller than the catalog value, and $\phase_0$ values slightly closer to the trailing side of the symmetric centre of the profile.

Given the finding that the fiducial point is likely not at the exact center of the profile, it is interesting to ask to what degree AR effects could contribute to the observed asymmetry.
If we assume either of the geometries proposed by \citet{Bhattacharyya2009}, and also assume that the outer emission cone is generated at (or near) the last open field lines, we find \citep[using the formulation of][]{Gangadhara2004a} that the emission heights are in the approximate range 200-300\,km.
Using this as a rough guide, we find that AR effects would not change the measured evolution rates by more than ${\sim}0.0005^\circ$/MHz.

Given the complexity of the PA curve, the use of the PA as a proxy for the angle $\psi$ may also be called into question, an assumption that relies heavily on the RVM being an appropriate model for \psr{}.
Just as being able to extend the above analysis to the inner component would be useful, so too would repeating the above analysis on data where the PA of individual pulses can be obtained, especially since the PAs of the inner component sub-pulses is much better behaved than that of the outer components \citep[see Fig. 5 of][]{Bhattacharyya2009}.

The results presented in this paper could be fruitfully complemented by analyses that exploit the average behavior of the drift bands, such as the aforementioned work of \citet{Bilous2018} or the multi-frequency analysis of \citet{Maan2019}.
As stated in the introduction, this may prove challenging for \psr{}, owing to the presence of nulls which frequently interrupt the otherwise coherent drifting behavior.
Nevertheless, if these difficulties can be overcome, investigating the frequency dependence of the inner components may prove vital in determining, for instance, whether the inner and outer beam carousels conjectured by \citet{Bhattacharyya2009} might be twin manifestations of a single carousel of sparks on the surface, or whether there is evidence for twisted magnetic field lines.

The application of our model to other pulsars with known sub-pulse modulation \citep[e.g. B1237+25;][]{Wang2022} is expected to shed further light on whether the carousel model is generally correct, and, if it is, what constraints are placed on the shape of the carousel.
However, performing this kind of analysis requires high S/N single pulses over a wide frequency range, which limits the number of pulsars that can be investigated in this way.
With recent advances in receiver technology, and with next-generation telescope coming online in the next five to ten years, more and more pulsars will be amenable to single pulse analyses, shedding further light on the pulsar beam geometries, and ultimately on the radio emission mechanism.

\begin{acknowledgements}

Part of this research has made use of the EPN Database of Pulsar Profiles maintained by the University of Manchester, available at: \url{http://www.jodrellbank.manchester.ac.uk/research/pulsar/Resources/epn/}.
LS was supported by an ICRAR Studentship and an EECMS Summer Scholarship through Curtin University.
The GMRT is run by the National Centre for Radio Astrophysics of the Tata Institute of Fundamental Research, India.
The authors thank S. Kudale for conducting the observations.
The scientific work also made use of Inyarrimanha Ilgari Bundara, the CSIRO Murchison Radio-astronomy Observatory.
We acknowledge the Wajarri Yamaji people as the traditional owners of the Observatory site.
GW thanks the University of Manchester for Visitor status.

\end{acknowledgements}

\bibliography{19SS-B0818-41}{}
\bibliographystyle{aasjournal}

\end{document}